\documentclass[]{spie}  

\usepackage{amsmath,amsfonts,amssymb}
\usepackage{graphicx}
\usepackage[colorlinks=true, allcolors=blue]{hyperref}

\usepackage{array, booktabs}
\usepackage{float}  
\usepackage{placeins} 
\usepackage{comment}
\usepackage{csquotes}
\usepackage{subcaption}
\usepackage{xcolor}

\def\equationautorefname~#1\null{Eq.~(#1)\null}

\title{SCOPE-MRI: Bankart Lesion Detection as a Case Study in Data Curation and Deep Learning for Challenging Diagnoses}

\author[1]{Sahil Sethi}
\author[1]{Sai Reddy}
\author[2]{Mansi Sakarvadia}
\author[3]{Jordan Serotte}
\author[3]{Darlington Nwaudo}
\author[3,*]{Nicholas Maassen}
\author[3,*]{Lewis Shi}
\affil[1]{Pritzker School of Medicine, University of Chicago, 
IL, USA
}
\affil[2]{Department of Computer Science, University of Chicago, 
IL, 
USA}
\affil[3]{Department of Orthopaedic Surgery \& Rehabilitation Medicine,
UChicago Medicine, 
IL, USA}

\authorinfo{\textit{*These authors contributed equally}\\
Author information: (Send correspondence to S.S.)\\S.S.: sethis@uchicago.edu\\  N.M.: nmaassen@bsd.uchicago.edu\\ L.S.: lshi@bsd.uchicago.edu}

\begin{document} 
\maketitle

\begin{abstract}

\noindent Deep learning has shown strong performance in musculoskeletal imaging, but prior work has largely targeted conditions where diagnosis is relatively straightforward. More challenging problems remain underexplored, such as detecting Bankart lesions (anterior-inferior glenoid labral tears) on standard MRIs. These lesions are difficult to diagnose due to subtle imaging features, often necessitating invasive MRI arthrograms (MRAs). We introduce ScopeMRI, the first publicly available, expert-annotated dataset for shoulder pathologies, and present a deep learning framework for Bankart lesion detection on both standard MRIs and MRAs. ScopeMRI contains shoulder MRIs from patients who underwent arthroscopy, providing ground-truth labels from intraoperative findings, the diagnostic gold standard. Separate models were trained for MRIs and MRAs using CNN- and transformer-based architectures, with predictions ensembled across multiple imaging planes. Our models achieved radiologist-level performance, with accuracy on standard MRIs surpassing radiologists interpreting MRAs. External validation on independent hospital data demonstrated initial generalizability across imaging protocols. By releasing ScopeMRI and a modular codebase for training and evaluation, we aim to accelerate research in musculoskeletal imaging and foster development of datasets and models that address clinically challenging diagnostic tasks. 
\end{abstract}


\section{Introduction}
\label{sec:intro}  
Deep learning (DL) in medical imaging has transformed healthcare research, improving non-invasive diagnosis and creating clinical decision aids across numerous specialties  \cite{barnett_case-based_2021, calli_deep_2021, sun_lesion-aware_2021, fritz_artificial_2022}. In musculoskeletal imaging, DL has achieved impressive results in diagnosing conditions such as anterior cruciate ligament (ACL) tears, meniscus injuries, and rotator cuff disorders \cite{zhang_deep_2020, rodriguez_artificial_2023, lin_deep_2023}. However, these studies target pathologies where clinicians already excel. For example, initial diagnosis of ACL tears rarely requires imaging because the Lachman test, a physical exam for ACL tears, achieves a sensitivity and specificity of 81\% \cite{van_eck_methods_2013}. On magnetic resonance imaging (MRI), sensitivity and specificity are as high as 92\% and 99\%, respectively \cite{smith_diagnostic_2016}. Similarly, radiologists routinely achieve high diagnostic accuracy for meniscus and rotator cuff tears \cite{rutten_detection_2010, bolog_reporting_2016}. This reduces the potential clinical impact of applying DL to the diagnosis of these pathologies.

In contrast, anterior-inferior glenoid labral tears, or Bankart lesions, represent a far more challenging diagnostic task. These lesions are a leading cause of shoulder instability, and frequently occur in patients with traumatic shoulder dislocations\cite{rutgers_recurrence_2022}. Without accurate diagnosis and timely treatment, Bankart lesions can lead to chronic instability, progressive bone loss, and significant reductions in quality of life \cite{kang_complications_2009, rutgers_recurrence_2022}. However, diagnosing these lesions on standard (non-contrast) MRIs is exceptionally challenging due to their subtle imaging features and frequent overlap with normal anatomy, resulting in high inter-observer variability, even among experienced radiologists \cite{loh_is_2016}.

\begin{figure}[t]
   \begin{center}
   \begin{tabular}{c} 
   \includegraphics[height=7.5cm]{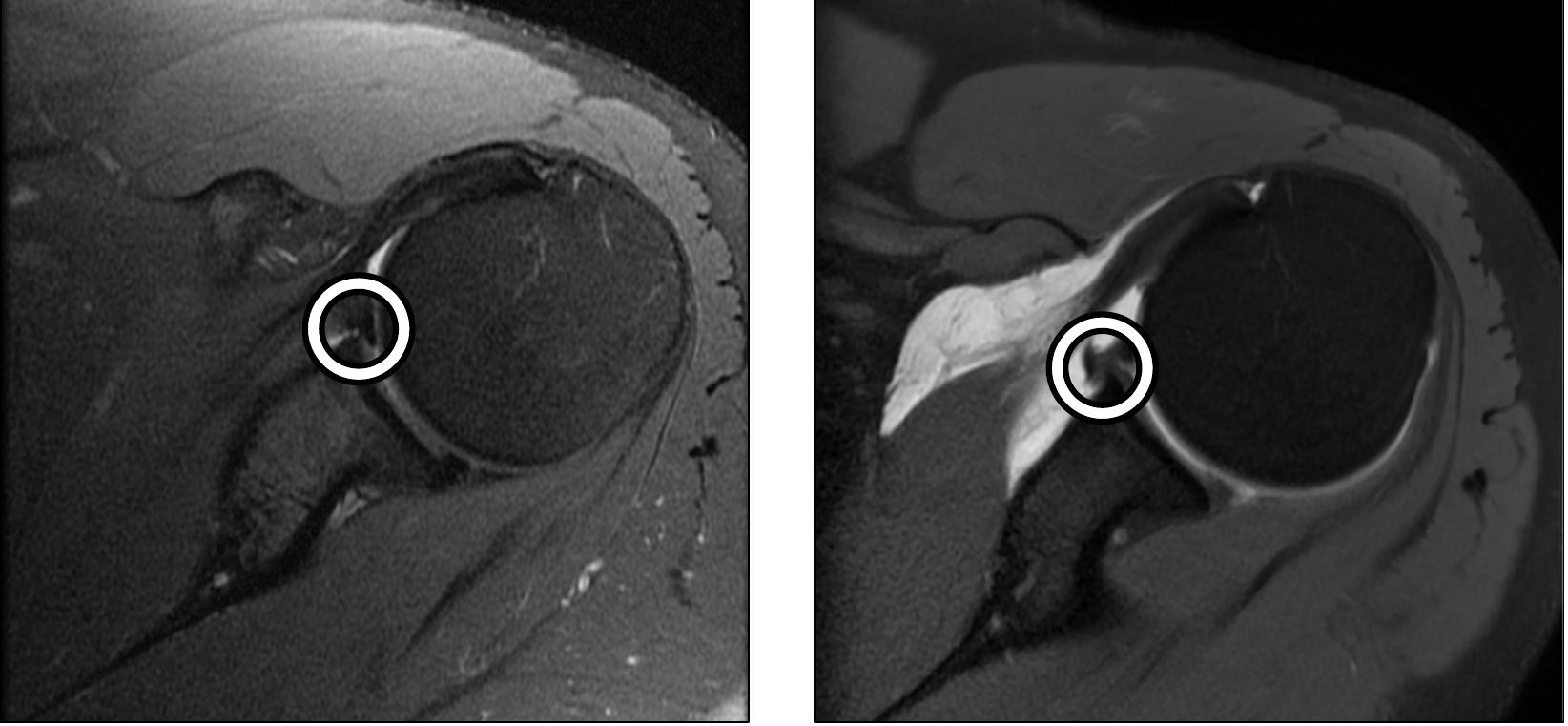}
   \end{tabular}
   \end{center}
   \caption[example] 
   { \label{fig:introfig} 
Bankart lesion on standard MRI (left) and MRI arthrogram (right) in the axial view. Images are from the same patient and depict the same tear. White circles reflect annotations identifying the tear,  provided by a shoulder/elbow fellowship-trained orthopedic surgeon.}
\end{figure}

Current clinical practice often relies on MRI arthrograms (MRAs), where contrast is injected into the joint to enhance visualization of labral structures \cite{chang_ssr_2024, chandnani_glenoid_1993}. The improved visualization of intra-articular structures is apparent in \autoref{fig:introfig}, which depicts the same Bankart lesion on an MRA and a standard (non-contrast) MRI. While MRAs achieve sensitivity and specificity rates of 74–96\% and 91–98\%, respectively \cite{rixey_accuracy_2023, magee_3-t_2009, woertler_mr_2006}, they are invasive, more expensive than standard MRIs, and associated with patient discomfort and risks such as allergic reactions and joint infections \cite{liu_comparison_2020, chang_ssr_2024, newberg_complications_1985, hugo_complications_1998, giaconi_morbidity_2011, ali_radio-carpal_2022}. Standard (non-contrast) MRIs, on the other hand, are non-invasive and more widely available but are significantly less reliable for detecting Bankart lesions, with reported sensitivity rates as low as 52–55\% \cite{arnold_non-contrast_2012, magee_sensitivity_2006}. This disparity underscores the urgent need for non-invasive diagnostic approaches that can match the diagnostic accuracy of MRAs while reducing patient burden and healthcare costs.

To address this unmet need, we introduce SCOPE-MRI (Shoulder Comprehensive Orthopedic Pathology Evaluation-MRI), the first publicly available, expert-annotated dataset for detecting Bankart lesions on shoulder MRIs. ScopeMRI provides a curated resource for advancing machine learning research in musculoskeletal imaging, a domain currently lacking accessible datasets. In addition to annotations for Bankart lesions, ScopeMRI includes expert labels for other clinically significant pathologies, such as superior labrum anterior-to-posterior (SLAP) tears, posterior labral tears, and rotator cuff tears. While this study focuses exclusively on Bankart lesions to demonstrate our approach, the inclusion of additional labels facilitates future research addressing diagnostic challenges in musculoskeletal imaging.

Our prior work demonstrated the initial feasibility of detecting Bankart lesions on standard MRIs using deep learning\cite{sethi_toward_2025}. Building on that preliminary work, this article introduces the ScopeMRI dataset, incorporates external validation data, includes a detailed comparison of model architectures and pretraining strategies, uses stratified cross-validation for stable architecture selection, and provides an analysis of model interpretability. These enhancements provide a comprehensive evaluation of our approach and its implications for tackling challenging diagnostic tasks in musculoskeletal imaging.

We present a comprehensive workflow demonstrating best practices for leveraging deep learning in small, imbalanced datasets, using Bankart lesions as a representative case study of a challenging diagnostic task. Our workflow includes several components. (1) Domain-Relevant Transfer Learning: we evaluated several state-of-the-art architectures on MRNet \cite{bien_deep-learning-assisted_2018}, including convolutional neural networks (CNNs) and transformers, and selected AlexNet \cite{krizhevsky_imagenet_2012}, Vision Transformer (ViT) \cite{dosovitskiy_image_2021}, and Swin Transformer V1 \cite{liu_swin_2021} for fine-tuning on our dataset; comparisons with ImageNet-initialized \cite{deng_imagenet_2009} models demonstrated superior performance with MRNet pretraining across all models and modalities, highlighting the value of domain-specific pretraining for specialized tasks (\ref{sec:model_selection}). (2) Stratified Cross-Validation for Stable Architecture Selection: we employed an eight-fold stratified cross-validation strategy to evaluate model performance variability across dataset partitions. The most stable architecture for each view-modality was identified and re-trained on the initial dataset split, ensuring a consistent validation set for training the multi-view ensemble (\ref{sec:model_stability}). (3) Multi-View Ensembling: by combining predictions across sagittal, axial, and coronal MRI views, we leveraged complementary diagnostic information to improve overall accuracy (\ref{sec:ensemble}). (4) External Validation and Interpretability: external testing on an independent dataset provided initial evidence of generalizability. Grad-CAM\cite{selvaraju_grad-cam_2017} visualizations highlighted alignment with clinically relevant features, offering insights into model decision-making (\ref{sec:interpretability}).

This study underscores the importance of curating and publicly releasing datasets for challenging diagnostic tasks, as widespread availability of high-quality datasets is essential for developing DL models that can generalize across diverse patient populations and imaging protocols. By demonstrating that even small, imbalanced datasets can yield clinically relevant performance when combined with effective DL methodologies, we highlight the feasibility and value of DL in low data regimes. Notably, our models achieved diagnostic performance on standard MRIs comparable to radiologists interpreting MRAs—a less accessible and more invasive imaging modality—demonstrating the impact of targeted dataset creation. With the release of ScopeMRI, we enable future research in musculoskeletal imaging and encourage broader contributions in this underexplored domain. To further support these efforts, we also release a publicly available code repository containing a general-purpose training pipeline for medical imaging, including cross-validation, hyperparameter tuning, and Grad-CAM visualization, which is readily adaptable to other binary classification tasks on MRI and computed tomography (CT) data.


\section{Results}
\label{sec:results}
\subsection{Model Selection \& Hyperparameter Tuning}
\label{sec:model_selection}

Results from the single best trial using the method described in \autoref{model_architecture} for each of the nine model types are presented in  \autoref{model_selection_extraresults_section}. AlexNet, Vision Transformer, and Swin Transformer V1 performed the best on the MRNet dataset\cite{bien_deep-learning-assisted_2018} (highest AUC), so were selected for fine-tuning on our dataset.  
 
\begin{figure}[t]
\centering
\includegraphics[height=12cm]{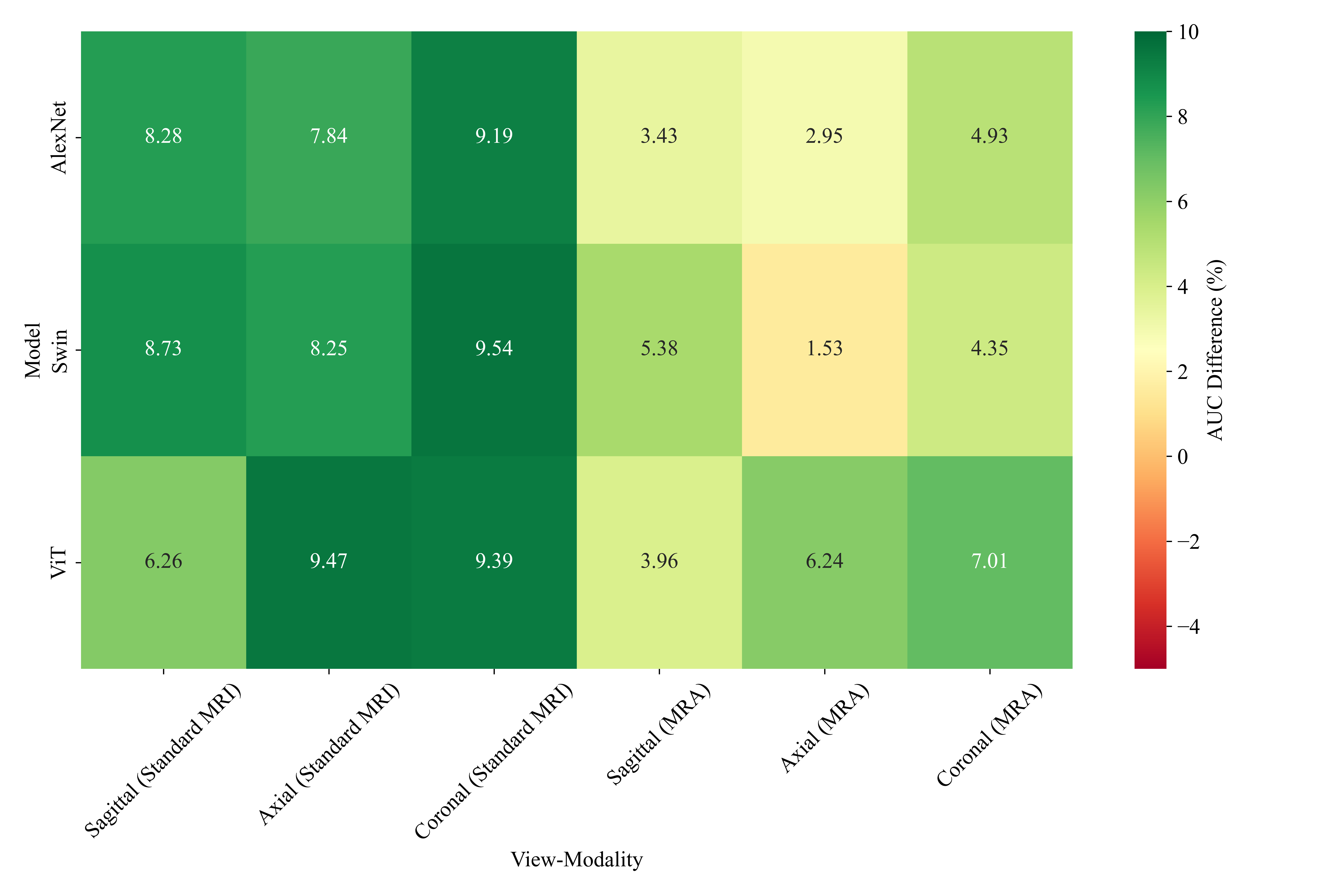}
\caption{
Heatmap illustrating the difference in validation AUC (MRNet - ImageNet) across model architectures (AlexNet, Swin Transformer, ViT) and view-modalities (sagittal, axial, coronal) for MRAs and standard MRIs. Positive values indicate higher performance with MRNet pretraining compared to ImageNet pretraining. Each cell represents the AUC difference for the corresponding model and view-modality pair, with results derived from the best-performing hyperparameter set for each model and view-modality. The AUC differences have been scaled by 100 for readability and are presented as percentages.
}
\label{fig:pretrain_auc_diff}
\end{figure}

To further optimize performance, we conducted 100 Hyperband\cite{li_hyperband_2018} hyperparameter tuning trials for each of the six view-modalities using the three model architectures (AlexNet, ViT, Swin Transformer). Each trial was initialized with MRNet-pretrained weights. We repeated this process with ImageNet initialization for comparison, resulting in 18 MRNet Hyperband loops and 18 ImageNet Hyperband loops—each loop consisting of 100 trials, with the \enquote{best} trial from each loop being the one achieving the highest validation AUC. The results of these \enquote{best} trials are presented in \autoref{fig:pretrain_auc_diff}, where we show the AUC difference between MRNet and ImageNet pretraining for each model type and view-modality.

\subsection{Model Stability Across Splits}
\label{sec:model_stability}

\begin{figure}[!tbh]
   \begin{center}
   \begin{tabular}{c} 
   \includegraphics[height=8cm]{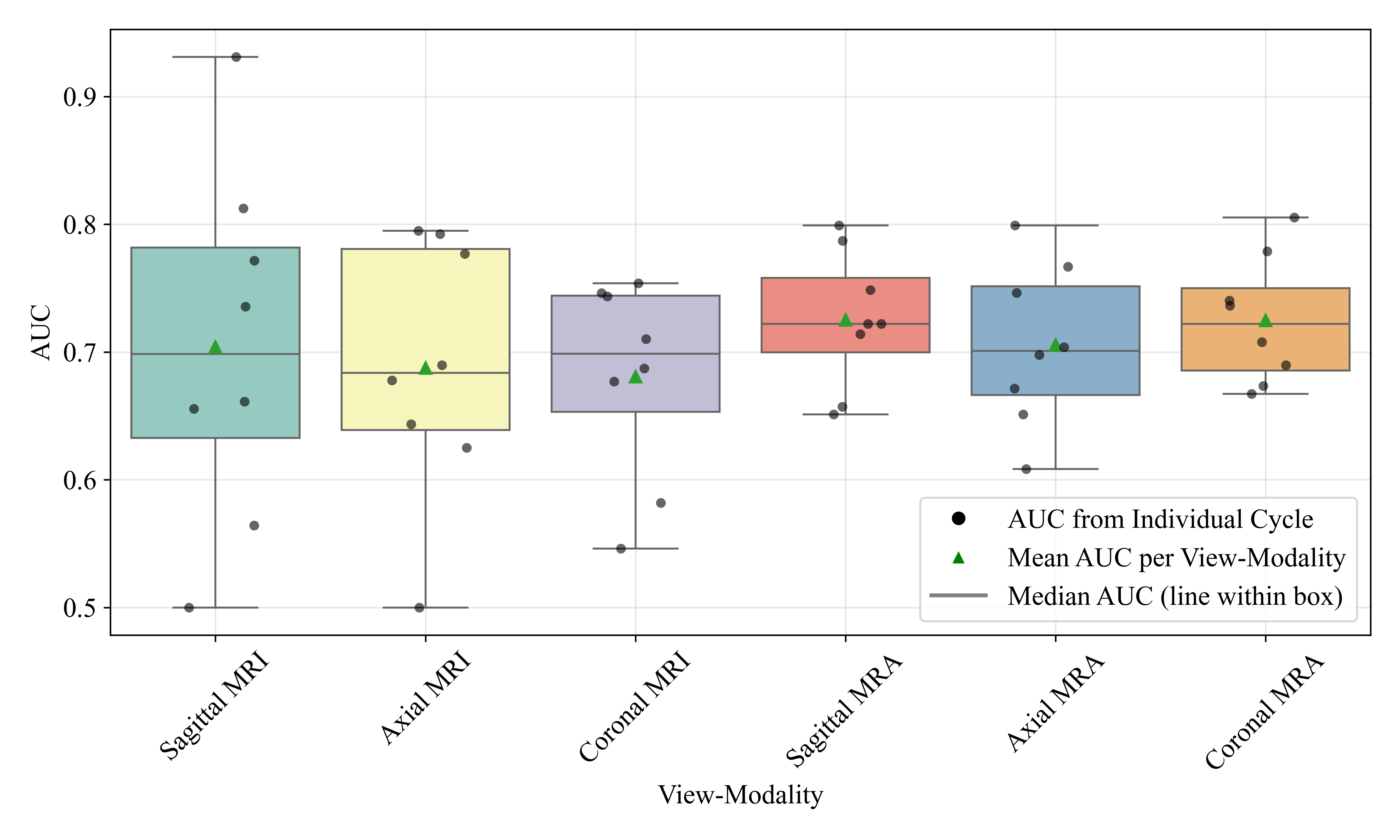}
   \end{tabular}
   \end{center}
   \caption {\label{fig:cross_validation} 
   Distribution of receiver operating characteristic (ROC) area under the curve (AUC) values across eight cross-validation splits for each view-modality's final selected architecture. ROC AUC quantifies the model’s ability to distinguish between classes. Each box shows the interquartile range (IQR, 25th–75th percentile), with whiskers extending to 1.5 times the IQR. The horizontal line within each box represents the median AUC, while green triangles indicate the mean AUC. Black dots depict individual split AUCs, with dots outside the whiskers representing outliers. This visualization demonstrates the model’s performance stability on different the validation folds across sagittal, axial, and coronal views for both standard MRIs and MRI arthrograms (MRAs).}
\end{figure}

\autoref{tab:cv_results} and \autoref{fig:cross_validation} show the cross-validation results for the models selected based on stability, defined by the lowest standard deviation in validation fold AUC. For Sagittal MRA, ViT was selected despite its lower mean AUC (0.725) compared to Swin Transformer (0.755), due to its substantially lower standard deviation (0.054 compared to 0.123). For the other view-modalities, the model with the lowest standard deviation of AUC also had the highest mean AUC, demonstrating that stability was generally aligned with performance. The full cross-validation results for the final models selected for each view-modality are presented in  \autoref{fig:cross_validation}. The hyperparameters for these models are provided in \autoref{hparam_tuning_extraresults_section}.

\begin{table}[t]
\centering
\caption{Cross-Validation Results for Each View-Modality. Cells contain AUC ± standard deviation; values for the final chosen architecture for each view-modality are bolded.}
\renewcommand{\arraystretch}{1.2}  
\begin{tabular}{p{0.15\linewidth}p{0.20\linewidth}p{0.15\linewidth}p{0.15\linewidth}p{0.20\linewidth}}
\toprule
\textbf{View-Modality} & \textbf{AlexNet (AUC)} & \textbf{Swin (AUC)} & \textbf{ViT (AUC)} & \textbf{Selected Model} \\
\midrule
Sagittal MRI   & 0.618 ± 0.179 & \textbf{0.704 ± 0.138} & 0.690 ± 0.187 & Swin Transformer \\
Axial MRI      & 0.668 ± 0.183 & 0.671 ± 0.223 & \textbf{0.688 ± 0.101}& ViT \\
Coronal MRI    & 0.663 ± 0.162 & \textbf{0.681 ± 0.078} & 0.658 ± 0.155& Swin Transformer \\
Sagittal MRA   & 0.720 ± 0.076 & 0.755 ± 0.123 & \textbf{0.725 ± 0.054} & ViT \\
Axial MRA      & \textbf{0.706 ± 0.063} & 0.705 ± 0.153 & 0.671 ± 0.101 & AlexNet \\
Coronal MRA    & 0.636 ± 0.109& 0.632 ± 0.172& \textbf{0.725 ± 0.050}& ViT \\
\bottomrule
\end{tabular}
\label{tab:cv_results}
\end{table}

\subsection{Multi-View Ensemble Performance}
\label{sec:ensemble}
The receiver operating characteristic (ROC) curves in \autoref{fig:ensemble_results}  depict the performance of the single-view models and multi-view ensemble on the hold-out test set. For the model to have predicted a Bankart lesion diagnosis, the multi-view ensemble's output probability must have exceeded the thresholds of 0.71 and 0.19 for standard MRIs and MRAs, respectively. These thresholds were chosen as described in \autoref{retrain_ensemble}. Results of the multi-view ensemble on the hold-out test set using these thresholds compared to radiologists are depicted in \autoref{tab:model_performance}. 

\begin{figure*}[t]
\centering

    \begin{subfigure}[b]{0.33\textwidth}
        \centering
        \includegraphics[width=\textwidth]
        {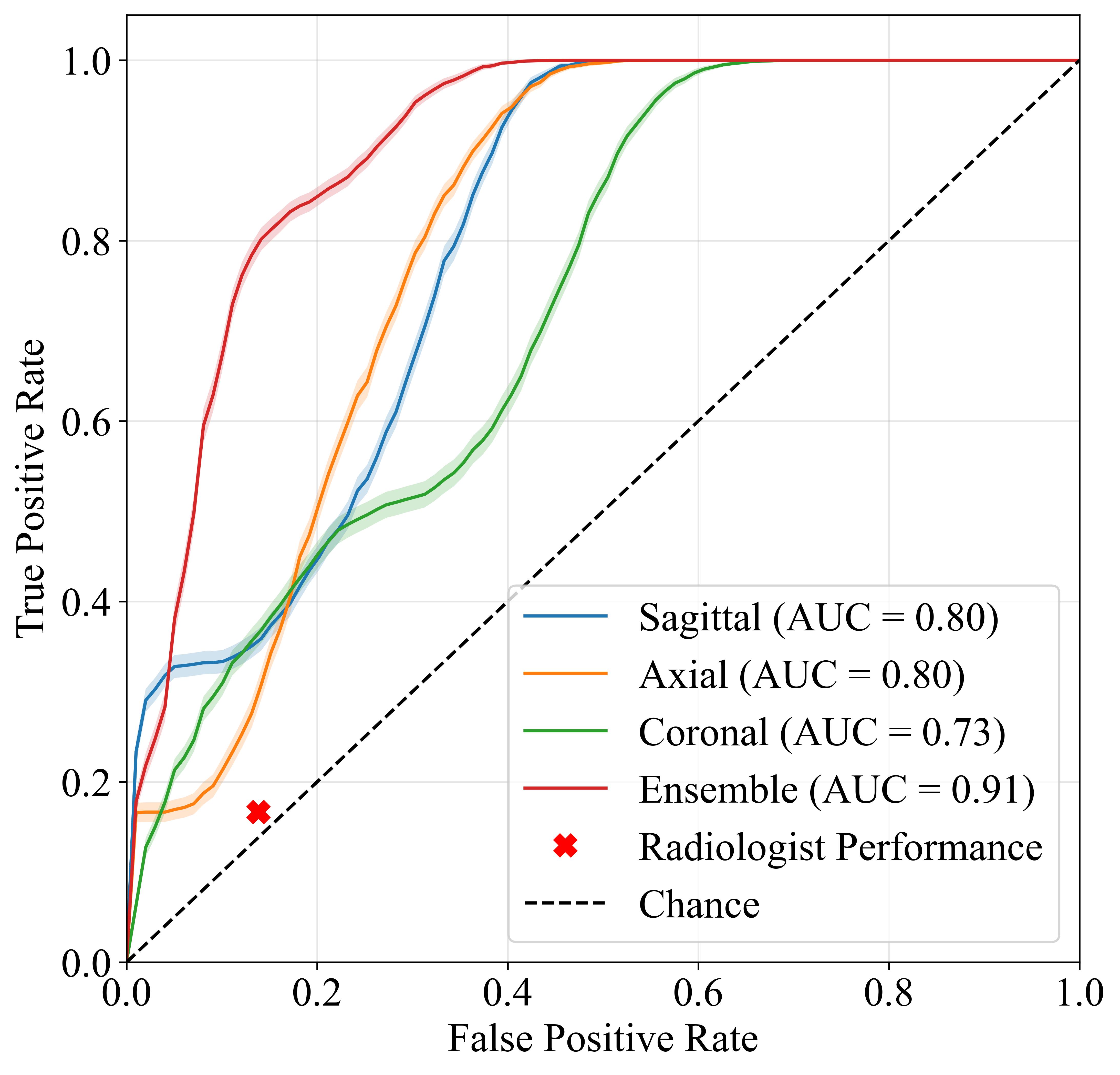}
        \caption{\textbf{Standard MRIs (Internal)}}
        \label{fig:roc_wo}
    \end{subfigure}%
    \hfill
    \begin{subfigure}[b]{0.33\textwidth}
        \centering
        \includegraphics[width=\textwidth]
        {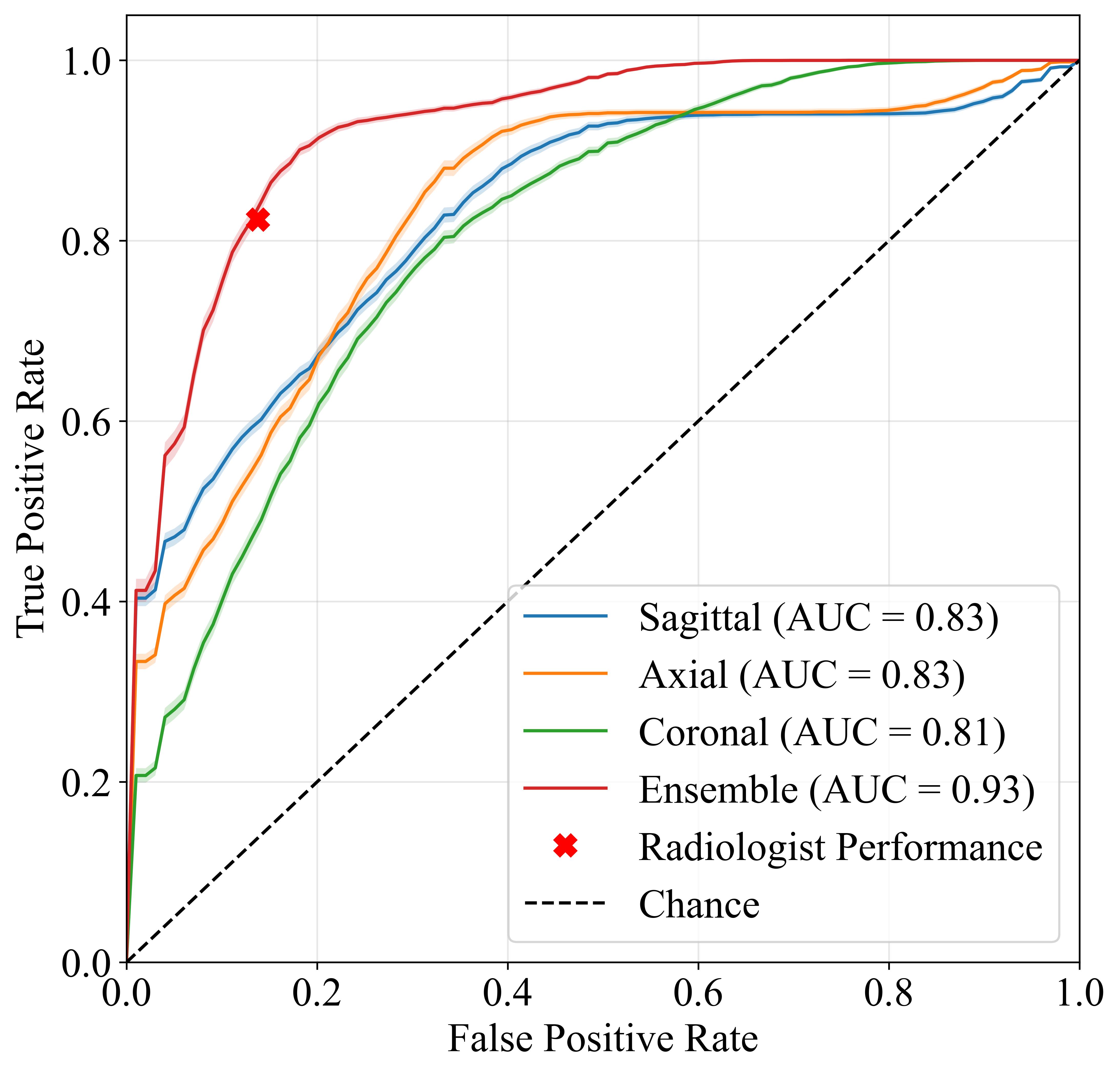}
        \caption{\textbf{MRAs (Internal)}}
        \label{fig:roc_w}
    \end{subfigure}%
    \hfill
    \begin{subfigure}[b]{0.33\textwidth}
        \centering
        \includegraphics[width=\textwidth]
        {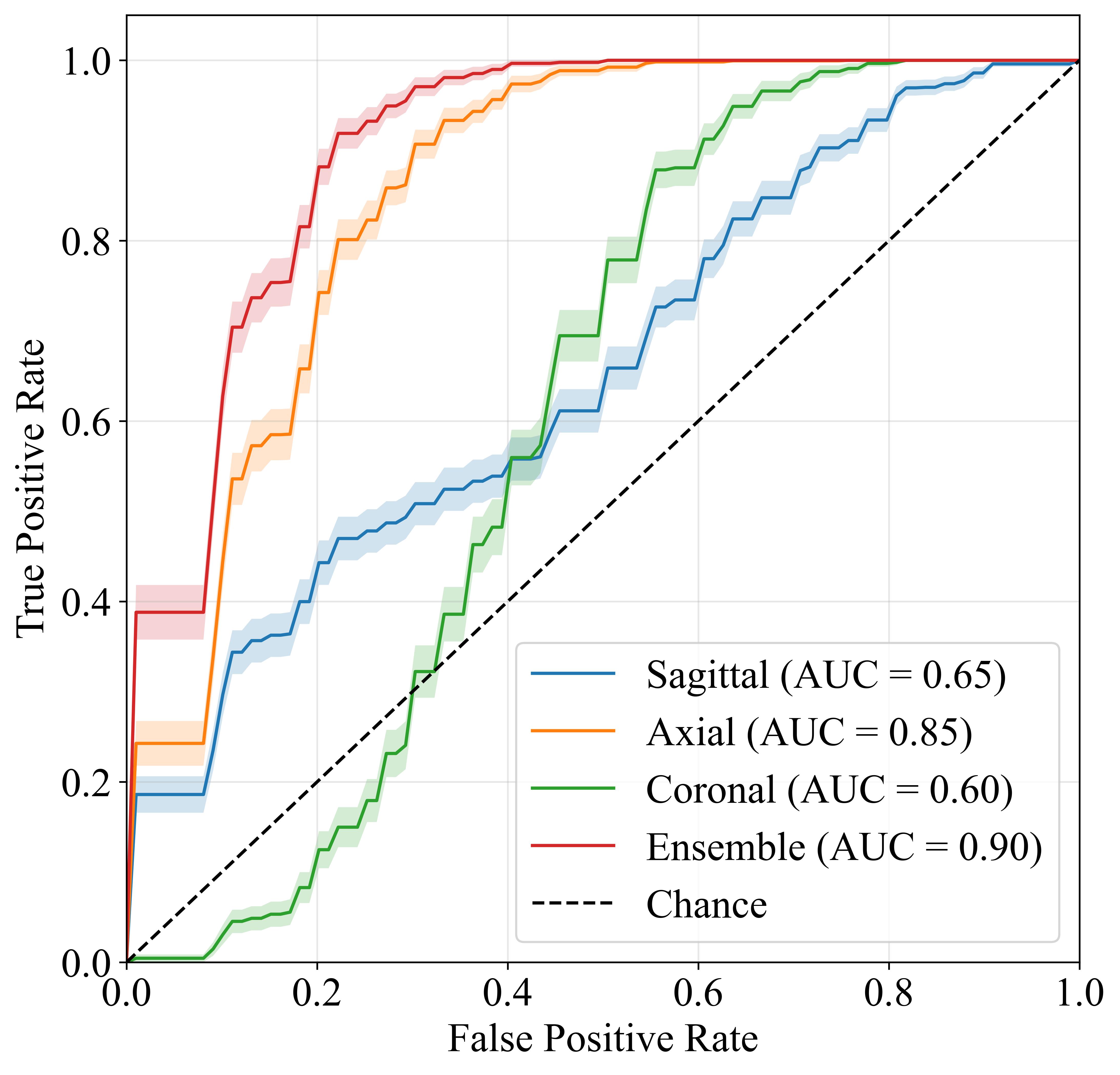}
        \caption{\textbf{Standard MRIs (External)}}
        \label{fig:roc_ext}
    \end{subfigure}%
  
    \caption{ \label{fig:ensemble_results} 
Receiver operating characteristic (ROC) curves for single-view models and the multi-view ensemble compared to radiologist performance. Results are shown for (a) internal standard MRIs, (b) MRI arthrograms (MRAs), and (c) external standard MRIs. The single-view models correspond to those included in the multi-view ensemble. Shaded regions around each curve represent 95\% confidence intervals, calculated through bootstrapping with 1000 iterations. Radiologist performance is marked with red X symbols, illustrating sensitivity and false positive rates derived from original radiology reports (internal datasets only). The dashed diagonal line indicates the performance of a random classifier (AUC = 0.50).
    }
    \label{fig:ensemble_results}
    
\end{figure*}

\begin{table}[t]
\centering
\caption{Final Ensemble Model Performance on Hold-Out Test Set \& External Dataset}
\renewcommand{\arraystretch}{1.2}  
\newcolumntype{P}[1]{>{\centering\arraybackslash}p{#1}}
\begin{tabular}{p{0.3\linewidth} P{0.15\linewidth} P{0.15\linewidth} P{0.15\linewidth} P{0.1\linewidth}}
\toprule
\textbf{} & \textbf{Accuracy} & \textbf{Sensitivity (Recall)} & \textbf{Specificity} & \textbf{AUC-ROC\textsuperscript{1}} \\
\midrule
\textbf{Test Standard MRIs} (n=71) & & & & \\ 
\emph{Model} & 90.14\% (64/71)& 83.33\% (5/6)& 90.77\% (59/65)& 0.9051 \\ 
\emph{Radiology Reports} & 80.28\% (57/71)& 16.67\% (1/6)& 86.15\% (56/65)&  -  \\ 
\emph{Literature Radiologists\textsuperscript{2}} & - & 52-55\% & 89-100\% & - \\ 

\midrule
\textbf{Test MRAs} (n=46) & & & & \\ 
\emph{Model} & 89.13\% (41/46)& 94.12\% (16/17)& 86.21\% (25/29)& 0.9256 \\ 
\emph{Radiology Reports} & 84.78\% (39/46)& 82.35\% (14/17)& 86.21\% (25/29)& -  \\ 
\emph{Literature Radiologists\textsuperscript{2}} & - & 74-96\% & 91-98\% & - \\

\midrule
\textbf{External Dataset} (n=12) & & & & \\ 
\emph{Model} & 83.33\% (10/12)& 100\% (2/2)& 80\% (8/10)& 0.9000 \\ 

\bottomrule
\end{tabular}
\vspace{5pt} 
\parbox{0.95\textwidth}{%
\footnotesize%
\textsuperscript{1} Area Under the Receiver Operating Curve (AUC-ROC).  \\
\textsuperscript{2} Values obtained for glenoid labral tears in general from largest studies available in the literature for standard MRIs  \cite{zlatkin_assessment_2004, arnold_non-contrast_2012} and MRAs \cite{rixey_accuracy_2023, magee_3-t_2009, woertler_mr_2006}. \\
}
\label{tab:model_performance}
\end{table}

\subsection{Model Interpretability}
\label{sec:interpretability}
\begin{figure}[t]
   \begin{center}
   \begin{tabular}{c} 
   \includegraphics[width=0.975\textwidth]{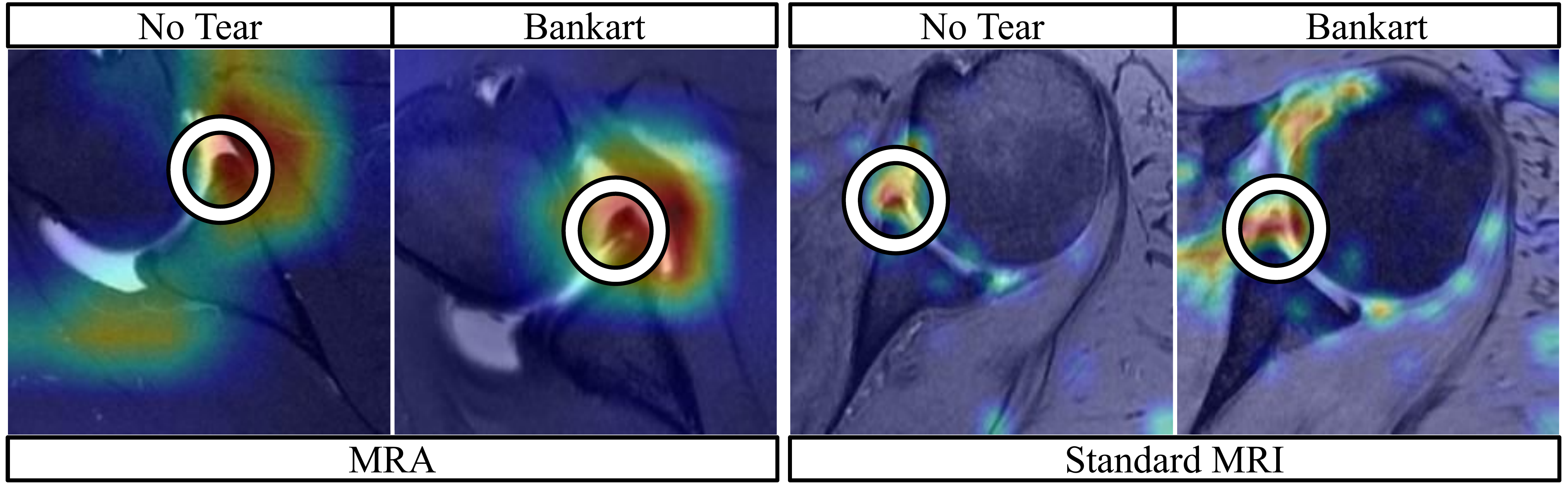}
   \end{tabular}
   \end{center}
   \caption[example] 
   { \label{fig:grad_cam} 
Gradient-weighted class activation mapping (Grad-CAM)\cite{selvaraju_grad-cam_2017} visualizations for Bankart lesion detection on MRAs (left) and standard MRIs (right) for the axial view. Cases with and without Bankart lesions are presented. The model correctly classified all four cases. White circles highlight the anterior labrum (the region of interest), annotated by a shoulder/elbow fellowship-trained orthopedic surgeon. Heatmaps indicate regions most influential to the model’s prediction, with warmer colors (red/yellow) signifying higher relevance.}
\end{figure}

Grad-CAM\cite{selvaraju_grad-cam_2017} visualizations highlight regions relevant to a model's predictions.  \autoref{fig:grad_cam} presents Grad-CAM heatmap visualizations for the axial view for four representative cases: an MRA without a tear, an MRA with a Bankart lesion, a standard MRI without a tear, and a standard MRI with a Bankart lesion. For all heatmaps, annotations highlight the anterior labrum, placed by a shoulder/elbow fellowship-trained orthopedic surgeon. The model attended to the relevant portions of the image, as determined by the surgeon, on all four cases. 

\section{Discussion}
\label{sec:discussion}
The heatmap in \autoref{fig:pretrain_auc_diff} reveals that all model-view-modalities showed positive performance improvements when MRNet pretraining was used compared to ImageNet initialization, confirming the benefit of using domain-specific pretraining across all models and view-modalities. Interestingly, when comparing the AUC improvements between standard MRIs and MRAs, we observed a significantly higher mean AUC difference (two-tailed unpaired p$<$0.001) for standard MRIs (mean = 8.55\%, sd = 1.06) compared to MRAs (mean = 4.42\%, sd = 1.69). This suggests that MRNet pretraining has a more substantial positive impact on performance for the more challenging, imbalanced standard MRIs than for MRAs, where contrast makes lesion detection is easier. These results further highlight the importance of domain-specific pretraining when working with challenging datasets.

The cross-validation results illustrate the variability in model performance across dataset splits. This variability underscores the need for careful model selection when working with small and imbalanced datasets. While we prioritized stability as a selection criterion, it is important to acknowledge that even the most stable models showed some variation in performance (\autoref{fig:cross_validation}). This underscores the importance of re-training on the initial random split (where hyperparameter tuning was performed) to avoid potential bias in selecting particularly performant splits from the cross-validation. Notably, the MRA results exhibited lower variability compared to standard MRI. This is expected, as the intra-articular contrast in MRAs generally enhances labrum visualization\cite{chang_ssr_2024}, possibly facilitating more stable and predictable model performance. For the easiest-to-interpret view-modality, the axial MRA\cite{knight_radiographic_2024}, the simplest architecture, AlexNet, was selected. In contrast, the Swin Transformer, being more complex than both AlexNet and ViT, was chosen for the two most challenging view-modalities—sagittal MRI and coronal MRI\cite{knight_radiographic_2024}. The Vision Transformer, which represents an intermediate level of complexity, was selected for the remaining view-modalities. This pattern suggests that simpler models perform more reliably on easier view-modalities, while more complex models are better suited for interpreting more challenging view-modalities.

The multi-view ensemble in \autoref{fig:ensemble_results} consistently outperformed single-view models across all datasets, demonstrating the value of integrating diagnostic information from sagittal, axial, and coronal views. Note that the single-view models in \autoref{fig:ensemble_results} performed better than during cross-validation (\autoref{fig:cross_validation}), likely due to differences in dataset size, tuning, and training schedules. The larger hold-out test set provided more stable estimates compared to the smaller validation folds, and hyperparameter tuning was specifically optimized for this dataset split. Additionally, the final re-training allowed for up to 100 epochs, compared to 30 epochs during cross-validation, potentially allowing the models to better learn Bankart lesion features. For standard MRIs (\autoref{tab:model_performance}), the ensemble achieved a sensitivity of 83.33\% and a specificity of 90.77\%, substantially exceeding the radiology reports on this dataset (16.67\% sensitivity, 86.15\% specificity). These results also compare favorably to the literature-reported sensitivity (52–55\%) and specificity (89–100\%) for radiologists interpreting standard MRIs \cite{arnold_non-contrast_2012, loh_is_2016}. The ensemble’s strong performance on standard MRIs addresses a critical limitation of this non-contrast imaging modality, which often struggles to reliably detect subtle Bankart lesions. Notably, the ensemble's sensitivity and specificity on standard, non-invasive MRIs compare to those typically reported for radiologists interpreting invasive MRAs (74–96\% sensitivity, 91–98\% specificity)\cite{rixey_accuracy_2023, magee_3-t_2009, woertler_mr_2006}, further underscoring its utility. Of note, it is expected that the radiology report performance on our standard MRI cohort is lower than published ranges. This is because our figures reflect routine clinical reports, whereas many literature estimates come from retrospective reader studies with consensus from multiple radiologists, settings that typically produce higher sensitivities than day-to-day reporting \cite{magee_sensitivity_2006}. The standard MRI test set also included only six Bankart-positive cases, making sensitivity volatile (one additional true positive would shift sensitivity by 16.7 percentage points). Together, the use of routine reports and the small number of positives plausibly explain the lower sensitivity observed here.

For MRAs (\autoref{tab:model_performance}), the ensemble achieved a sensitivity of 94.12\% and a specificity of 86.21\%, exceeding the sensitivity and matching the specificity of radiology reports on this dataset (82.35\% sensitivity, 86.21\% specificity). These results align well with literature benchmarks for radiologists interpreting MRAs, which report sensitivity and specificity ranges of 74–96\% and 91–98\%, respectively \cite{rixey_accuracy_2023, magee_3-t_2009, woertler_mr_2006}. The ensemble’s ability to achieve performance comparable to radiologists interpreting MRAs reinforces its ability to effectively combine diagnostic information across views and leverage the enhanced contrast provided by MRAs. Moreover, this performance on the less-imbalanced MRA test set (37\% positive cases) increases confidence in these results, as the MRAs provide a more balanced evaluation compared to the standard MRI dataset. 

The external dataset results in \autoref{fig:ensemble_results} and \autoref{tab:model_performance} provide preliminary evidence of generalizability, with the ensemble achieving an AUC of 0.90, perfect sensitivity (100\%), and a specificity of 80\%. These results suggest that the model's performance is not overly dependent on the internal dataset's characteristics. However, the small size of the external dataset (n = 12) and the low number of positive cases (n = 2) underscore the need for further validation on larger, more diverse datasets to confirm broader applicability. Overall, the ensemble’s performance highlights its ability to address the diagnostic challenges associated with Bankart lesions, a subtle pathology that often eludes detection on non-contrast imaging. By achieving sensitivity and specificity comparable to radiologists interpreting MRAs, the ensemble offers a potential non-invasive alternative to MRAs, particularly in settings where invasive imaging is less accessible or contraindicated. Further research is needed to validate its performance across diverse clinical settings and patient populations. 

The Grad-CAM visualizations in \autoref{fig:grad_cam} for the MRAs demonstrate focused activation within the anterior labrum for both the no-tear case and Bankart lesion case. For the no-tear MRA, the area of highest activation is in the anterior-inferior labrum region, but there is an additional lesser area of activation near the posterior labrum. In the Bankart lesion case, there is strong overlap between the activation and the surgeon-annotated region highlighting the tear, likely indicating that the model successfully identified the pathological features associated with the tear. For the standard MRIs, the Grad-CAM visualizations show less localized activation compared to MRAs, with several peripheral areas of minor activation near the image edges in both cases. This reduced specificity in the activation maps may result from the lower visual clarity of labral structures on standard MRIs due to the lack of intra-articular contrast. Additionally, as noted in \autoref{tab:demographics}, the standard MRIs had a significantly lower percentage of  3.0T exams (i.e. higher proportion of lower quality 1.5T exams)—suggesting that the lower average image quality in the standard MRI group may also have contributed to the less focused heatmap. Although the standard MRI dataset was larger overall (335 MRIs vs. 251 MRAs), it contained far fewer positive cases (8.6\% vs. 31.9\%), creating substantial class imbalance. This imbalance likely limited the model’s ability to learn localized features, leading to more diffuse activations. The broader activations likely reflect a combination of these factors, underscoring the need for cautious interpretation. For the no-tear standard MRI case, the heatmap shows the highest activation in the anterior labrum, suggesting that the model still prioritizes this region as a key diagnostic feature, even in the absence of pathology. For the Bankart lesion case, the anterior labrum remains the area of highest activation, but the heatmap also strongly highlights additional regions within the joint space. Unlike for the MRA, the annotating surgeon noted no obvious tear characteristics for the standard MRI with a Bankart lesion. Consequently, the additional highlighted regions in the model's heatmap may reflect subtle imaging cues associated with the tear that are not easily discernible to human observers. This suggests that the model may capture features indicative of the tear that are challenging for radiologists or surgeons to detect on standard MRIs due to the lack of intra-articular contrast.

Recent advancements in deep learning (DL) have significantly enhanced the detection of musculoskeletal pathologies in MRI scans. Notably, DL models have been developed for diagnosing anterior cruciate ligament (ACL) tears, meniscus injuries, and rotator cuff disorders \cite{fritz_artificial_2022, rodriguez_artificial_2023, lin_deep_2023}. For example, Zhang et al.\cite{zhang_deep_2020} applied DL algorithms to ACL tear detection, achieving diagnostic performance comparable to that of experienced radiologists. However, like many studies in musculoskeletal imaging, the data used in this work is not publicly available. While such studies demonstrate impressive results, they largely focus on pathologies where clinicians already excel. For instance, the Lachman test—a physical exam for detecting complete ACL ruptures—has a sensitivity and specificity of 81\% \cite{fritz_artificial_2022, van_eck_methods_2013}, and imaging is often used for confirmation and surgical planning rather than for primary diagnosis. Similarly, MRI-based models for meniscus and rotator cuff tears address diagnostic tasks where radiologists already achieve high performance \cite{rutten_detection_2010, bolog_reporting_2016}, limiting their clinical impact. In contrast, the detection of subtle musculoskeletal injuries, such as Bankart lesions, on MRI presents significant challenges, particularly in the absence of intra-articular contrast \cite{fallahi_indirect_2013}. A recent survey of publicly available MRI datasets aimed to comprehensively catalog existing resources, but notably, no shoulder datasets were identified, underscoring a critical gap in the field \cite{dishner_survey_2024}. In contrast, datasets like MRNet \cite{bien_deep-learning-assisted_2018} and fastMRI \cite{zbontar_fastmri_2019} have catalyzed progress in knee and other anatomical regions by providing resources for reproducibility and benchmarking. The lack of comparable datasets for shoulder pathologies severely limits the development of DL models targeting several clinically challenging diagnoses. Existing studies on labral tears have primarily focused on superior labrum anterior-to-posterior (SLAP) tears. Ni et al.\cite{ni_deep_2022} used a convolutional neural network (CNN) to detect SLAP tears on MRI arthrograms (MRAs), which improve diagnostic clarity but are invasive, costly, and associated with procedural risks \cite{chang_ssr_2024, ali_radio-carpal_2022, newberg_complications_1985, hugo_complications_1998}. Clymer et al., in contrast, examined SLAP tears on standard MRIs but focused on pretraining methodologies rather than addressing broader clinical challenges \cite{clymer_applying_2019}. Importantly, no prior studies have targeted Bankart lesion detection—which, on standard MRIs, is a task that radiologists themselves find exceedingly difficult, achieving only 16.7\% sensitivity in our dataset. Our study is the first to address this unmet clinical need by developing DL models for Bankart lesion detection on both standard MRIs and MRAs. Unlike prior studies, we use all available MRI sequences (e.g., T1, T2, MERGE, PD, STIR) across sagittal, axial, and coronal views, and explored state-of-the-art architectures, including both CNNs and transformers. By leveraging domain-specific pretraining, stratified cross-validation, and multi-view ensembling, our models achieve diagnostic performance on non-invasive standard MRIs that rivals radiologist performance on MRAs—addressing a longstanding clinical challenge. Additionally, we introduce ScopeMRI, the first publicly available, expert-annotated shoulder MRI dataset. ScopeMRI includes image-level annotations for Bankart lesions, SLAP tears, posterior labral tears, and rotator cuff injuries, providing a critical resource for advancing DL applications in shoulder imaging. By addressing both a clinically significant unmet need and the systemic lack of public datasets, our work demonstrates the potential of DL to improve diagnostic workflows for challenging diagnoses in musculoskeletal imaging.

Several limitations of this study should be noted. First is the small number of positive cases in the test set (6/71, 8.45\%) for standard MRIs. This imbalance reflects current clinical practice, where standard MRIs are rarely ordered when a Bankart lesion is suspected due to their perceived lower diagnostic utility compared to MRAs. While our findings demonstrate the potential of deep learning to improve the utility of standard MRIs, the limited sample size and class imbalance constrained statistical power and may reduce the stability of reported performance metrics. Larger, multi-institutional datasets will be required for future validation. By publicly releasing ScopeMRI, our goal is to help enable such studies. Additionally, the external dataset of 12 standard MRIs provides an initial effort to assess generalizability by introducing variability in imaging protocols, including differences in manufacturers and acquisition settings. However, this dataset does not represent a distinct patient population, as all included patients underwent surgery at our institution. Further, all 12 scans were standard MRIs, so the generalizability of our MRA models to external scans remains untested. Despite these limitations, incorporating variability in standard MRI imaging protocols is an important step toward understanding how models might perform across diverse imaging conditions. Finally, while our cross-validation approach was designed to identify the most stable architectures, our results still demonstrated performance variability across splits. This variability reflects the inherent challenges of small datasets, and results derived from such datasets should always be interpreted with caution. 

A future reader/observer study is critical to evaluate the model’s impact on clinical decision-making. While our models outperformed radiologists' original reports on our dataset, they should be used as decision-support tools rather than standalone decision-makers \cite{barnett_case-based_2021, fritz_artificial_2022}. A reader study would assess how clinicians interact with the model’s outputs and compare clinician model-assisted and model-unassisted performance, providing insights into the practical utility of the models and their potential to enhance diagnostic confidence and accuracy.

Enhancing model interpretability remains a critical step toward clinical adoption. This study used Grad-CAM \cite{selvaraju_grad-cam_2017} as a post-hoc interpretability tool to visualize the regions most contributing to the model's predictions. While Grad-CAM provides useful insights, it has limitations, including variability based on chosen model layer and input characteristics \cite{omeiza_smooth_2019, buono_expected_2024, lucas_visual_2022}. Given these challenges, Grad-CAM is best viewed as a tool for approximate understanding rather than a definitive validation mechanism. Future work should explore integrating interpretability techniques directly into the model architecture, such as the prototype-based \enquote{This Looks Like That} framework \cite{chen_this_2019}, which enables models to reason through clinically interpretable features. This approach has shown promise in other medical domains, such as breast cancer \cite{barnett_case-based_2021}, electroencephalography \cite{barnett_improving_2024}, and electrocardiography \cite{sethi_protoecgnet_2025-1}, and could improve trust and usability among clinicians by aligning model reasoning with familiar diagnostic patterns.

A transformative direction for this research lies in exploring techniques to convert standard MRIs into MRA-like images. This would be similar to the \enquote{virtual staining} approach that has shown success in digital histopathology \cite{latonen_virtual_2024}, and could enhance interpretability and usability for radiologists by creating synthetic MRA-like images that preserve the diagnostic familiarity of traditional arthrograms. However, this would be most feasible with paired standard MRI-MRA datasets, which may be challenging to acquire. Future work should investigate the feasibility of generating such datasets and the performance of models trained on these synthetic images.

We propose a deep learning workflow that successfully detects Bankart lesions on both standard MRIs and MRAs, achieving high accuracy, sensitivity, and specificity on both modalities. Our models address a critical clinical need by enabling diagnostic performance on non-invasive standard MRIs comparable to radiologists interpreting MRAs, a more invasive, more costly, and less accessible modality. This approach has the potential to reduce reliance on arthrograms, minimizing patient burden and enhancing diagnostic accessibility, particularly in resource-limited settings. Alongside addressing this diagnostic challenge, we introduce ScopeMRI, the first publicly available, expert-annotated shoulder MRI dataset. By providing annotations for multiple clinically significant shoulder pathologies, ScopeMRI supports future research and facilitates the development of diagnostic models for complex musculoskeletal conditions. Our study demonstrates that even small, imbalanced datasets can achieve clinically relevant results if effectively used, highlighting the value of similar dataset creation efforts for other diagnostic challenges. To enable reproducibility and broader application, we also release a modular deep learning codebase designed for binary classification on MRIs and CTs. While developed for Bankart lesion detection, the framework supports training, cross-validation, hyperparameter tuning, and Grad-CAM visualization, and is readily adaptable to other diagnostic tasks and imaging modalities. Future work should validate our findings on larger, external datasets to confirm their generalizability across diverse clinical environments. Furthermore, integrating interpretability methods into diagnostic workflows will be essential for fostering clinical trust and adoption. These efforts are critical for translating technological advancements into meaningful improvements in patient care.

\section{Methods}
\label{sec:methods}
\subsection{SCOPE-MRI Dataset Collection}

\begin{figure}[t]
   \begin{center}
   \begin{tabular}{c} 
   \includegraphics[width=\textwidth]{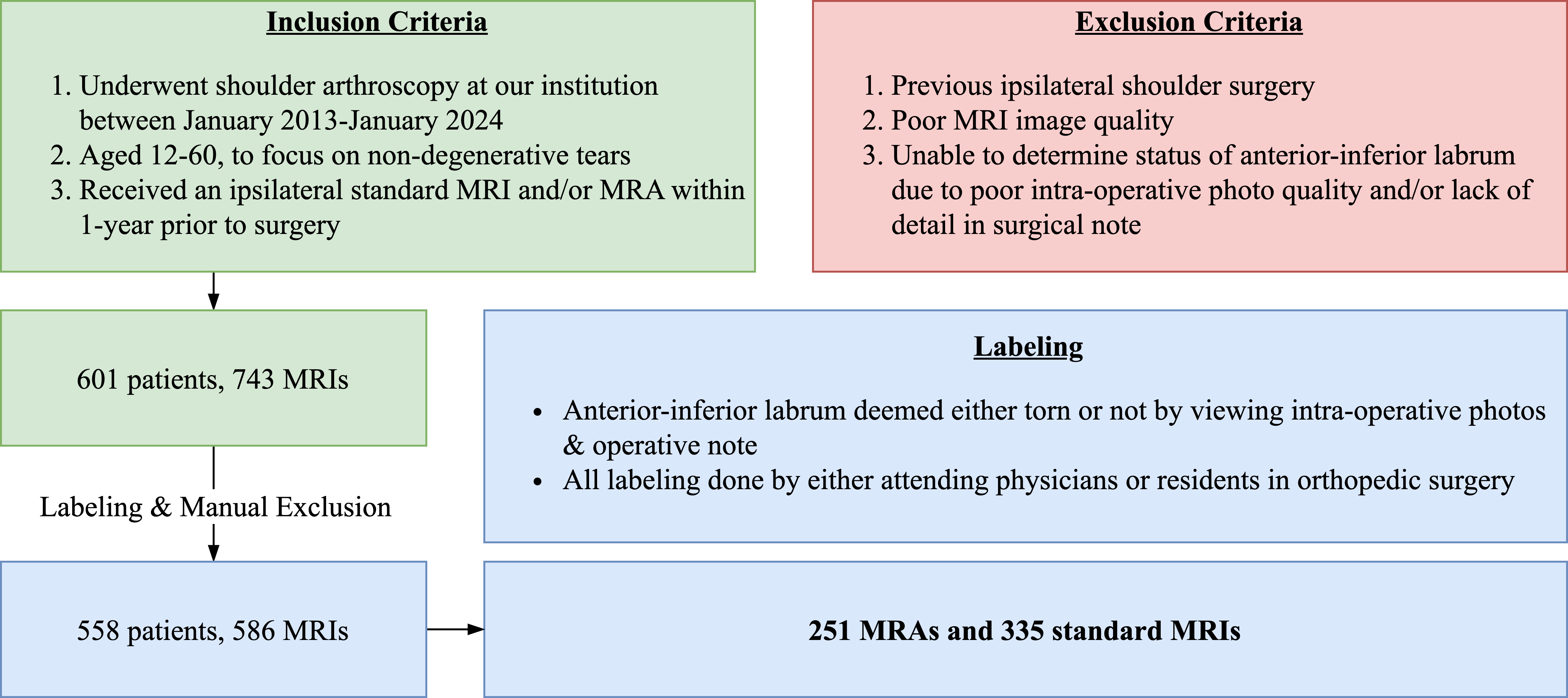}
   \end{tabular}
   \end{center}
   \caption[example] 
   { \label{fig:data_collection} 
Data Collection and Labeling Protocol.}
\end{figure}

\autoref{fig:data_collection} overviews the data collection and labeling protocol. The ScopeMRI dataset consists of patients who underwent shoulder arthroscopy for any diagnosis at University of Chicago Medicine between January 2013 and January 2024. Patients aged 12 to 60 who had received a standard shoulder MRI and/or MRI arthrogram (MRA) within one year prior to arthroscopic surgery of the ipsilateral shoulder were included. Children under the age of 12 typically have open physes and therefore different pathology of injury in the shoulder. We excluded patients greater than 60 years old as our study focused on acute, potentially operative labral tears, and older patients typically have degenerative labral tears. University of Chicago Medicine's clinical research database was queried to identify medical record numbers (MRNs) and clinical information for patients meeting these criteria. This information included accession numbers for the appropriate pre-operative standard MRI/MRA for each patient. University of Chicago Medicine's research imaging office then de-identified the scans and provided them along with a unique identifier key. This process yielded 601 patients with 743 MRIs. Additional exclusion criteria included patients with prior shoulder surgery on the ipsilateral shoulder, insufficient intraoperative photos, or operative notes with insufficient detail. Manual exclusion was performed for these criteria through chart review, narrowing the dataset down to 546 patients with 586 MRIs (335 standard MRIs and 251 MRAs). Labels were obtained during this manual chart review by accessing intraoperative arthroscopy photos alongside the operative note associated with the surgery following each standard MRI or MRA. The anterior-inferior labrum was then classified as either torn (Bankart lesion) or intact based on this information. Binary labels were also created for SLAP tears, posterior labral tears, and rotator cuff tears (i.e. any type of rotator cuff tear vs. no tear). Labels were curated by two shoulder/elbow fellowship-trained orthopedic surgeons and two orthopedic surgery residents trained by the surgeons. A common subset of 20 MRIs was labeled by all raters to measure inter-rater reliability—they achieved a Fleiss’s kappa=1.0 for all four labels included in ScopeMRI, indicating complete agreement.

\subsection{SCOPE-MRI Dataset Characteristics}
After applying exclusion criteria (detailed in \autoref{fig:data_collection}), our final dataset was narrowed down to 558 patients and 586 MRAs/standard MRIs (586 shoulders; some patients had bilateral surgeries). Dataset demographics and characteristics are in \autoref{tab:demographics}. 

Out of the 586 MRIs, there were 109 (18.5\%) with Bankart lesions based on intraoperative photos in correlation with the surgeon's operative note (diagnostic gold standard). This ratio was approximately the same in the training, validation, and testing sets due to using random stratified sampling. The MRA and standard MRI groups varied significantly. Of the 251 MRI arthrograms, 80 (31.9\%) had Bankart lesions, while only  8.6\% (29/335) of standard MRIs had Bankart lesions (p $<$ 0.001). MRA patients were significantly younger than standard MRI patients (31.0 versus 48.6 years old, p $<$ 0.001). Moreover, there was a lower percentage of female patients in the MRA group than the standard MRI group (35.1\% versus 50.1\%; p $<$ 0.001). The percent of right-sided exams between both groups was similar (61.0\% MRA versus 63.3\% standard MRI; p=0.565). However, there was a higher proportion of 3.0T scans for MRAs than for standard MRIs (73.3\% versus 54.6\%; p $<$ 0.001). The remaining exams were all 1.5T; all were taken with either a GE Medical Systems or Philips Healthcare machine. 

The differences in the demographic and clinical characteristics between the MRA and standard MRI cohorts are consistent with current clinical practices. Younger patients, who are more likely to be suspected of having labral tears, are often preferentially given MRAs due to their demographic and activity levels. Consequently, the higher prevalence of labral tears in the MRA group aligns with the expectation that MRAs are ordered when there is a stronger suspicion of labral pathology \cite{woertler_mr_2006}. 

\begin{table}[b]
\centering
\caption{SCOPE-MRI Demographics and Characteristics}
\renewcommand{\arraystretch}{1.2}  
\newcolumntype{P}[1]{>{\centering\arraybackslash}p{#1}}
\begin{tabular}{p{0.35\linewidth}  P{0.12\linewidth}  P{0.12\linewidth}  P{0.12\linewidth} P{0.12\linewidth}}
\toprule
\textbf{Characteristic} & \textbf{Total MRIs\textsuperscript{1}} & \textbf{MRI Arthrograms} & \textbf{Non-Enhanced MRIs} & \textbf{p-value\textsuperscript{2}}  \\
\midrule
\textbf{} & \textbf{n = 586} & \textbf{n = 251} & \textbf{n = 335} & \\ 
\midrule
\textbf{Number of Patients\textsuperscript{3}} & 546 & 238 & 318 & - \\ 
\textbf{Average Age (SD)} & 41 (13.9) & 31 (12.7) & 48.6 (9.3) & p $<$ 0.001 \\ 
\textbf{Female Sex (\%)} & 256 (43.7) & 88 (35.1) & 168 (50.1) & p $<$ 0.001\\ 
\textbf{Right-Sided Exams (\%)} & 365 (62.3) & 153 (61.0) & 212 (63.3) & p = 0.565\\ 
\textbf{3.0T Exams (\%)} & 367 (62.6) & 184 (73.3) & 183 (54.6) & p $<$ 0.001\\ 
\midrule
\textbf{\# MRIs with Bankart Lesions (\%)} & 109 (18.5) & 80 (31.9) & 29 (8.6) & p $<$ 0.001\\ 
\midrule
\textbf{Initial Dataset Split} & & & & \\ 
Training MRIs (\% with tear) & 410 (18.5) & 186 (30.1) & 224 (8.7) & - \\ 
Validation MRIs (\% with tear) & 59 (18.6) & 19 (36.8) & 40 (10.0) & - \\ 
Testing MRIs (\% with tear) & 117 (18.8) & 46 (37.0) & 71 (7.0) & - \\ 
\bottomrule
\end{tabular}
\vspace{5pt} 
\parbox{0.95\textwidth}{%
\footnotesize%
\textsuperscript{1} All available sequences were included for each MRI (e.g., T1, T2, MERGE, PD, STIR), yielding 1109 axial, 1647 coronal, 978 sagittal, and 237 ABER (Abduction and External Rotation) sequences; ABER sequences were excluded due to insufficient numbers for model training. \\
\textsuperscript{2} p-values obtained via chi-squared for categorical variables and unpaired two-tailed t-test for continuous variables. \\
\textsuperscript{3} Total does not add up as some patients had surgeries on both sides, and may have received an MRA on one side and a standard MRI on the other. \\

}
\label{tab:demographics}
\end{table}

\subsection{External Dataset Collection \& Description:}
In addition to the ScopeMRI dataset, we included an external dataset to evaluate the model’s ability to generalize to imaging protocols from other institutions. External dataset demographics and characteristics are displayed in \autoref{tab:ext_demographics}. This external dataset consisted of 12 standard shoulder MRIs performed at outside institutions; no MRAs were available. These standard MRIs were imported into University of Chicago Medicine’s records because the patients subsequently underwent arthroscopic surgery at our institution. Ground truth labels for these MRIs were derived using the same method as for the ScopeMRI dataset: findings from arthroscopy photos and documented in the surgical notes. The external dataset differs from the ScopeMRI dataset in imaging protocols, including variations in MRI machine manufacturers, magnet strengths, and acquisition settings. Specifically, it includes four 3.0T, seven 1.5T, and one 1.0T scan; manufacturer information is as follows: eight Siemens, one Philips Healthcare, one GE Health Systems, and two Hitachi machines.  

\begin{table}[b]
\centering
\caption{External Dataset Demographics and Characteristics}
\renewcommand{\arraystretch}{1.2}  
\newcolumntype{P}[1]{>{\centering\arraybackslash}p{#1}}
\begin{tabular}{p{0.35\linewidth}  l }
\toprule
\textbf{Characteristic} & \textbf{Non-Enhanced MRIs\textsuperscript{1}} \\
\midrule
\textbf{} & \textbf{n = 12}\\ 
\midrule
\textbf{Number of Patients} & 12 \\ 
\textbf{Average Age (SD)} & 35.8 (13.4) \\ 
\textbf{Female Sex (\%)} & 8 (66.7) \\ 
\textbf{Right-Sided Exams (\%)} & 9 (75.0) \\ 
\textbf{3.0T Exams(\%)} & 4 (33.3) \\ 
\midrule
\textbf{\# MRIs with Bankart Lesions (\%)} & 2 (16.7) \\ 
\bottomrule
\end{tabular}
\vspace{5pt} 
\parbox{0.95\textwidth}{%
\footnotesize%
\textsuperscript{1} All available sequences were included for each MRI (e.g., T1, T2, MERGE, PD, STIR), yielding 19 axial, 30 coronal, 27 sagittal sequences.
}
\label{tab:ext_demographics}
\end{table}

\subsection{Image Preprocessing}
Imaging data was preprocessed to prepare it for our deep learning study (see \autoref{sec:methods}). Preprocessing involved resizing the volumes to a shape of $n$ x 400 x 400 pixels, where $n$ was the number of slices in the MRI. Then, the slices were center-cropped to 224 x 224 to isolate the region of interest. Standardization was performed for exams of each series/sequence type—T1, T2, MERGE, PD, or STIR (separately for with vs. without fat-saturation, as applicable)—after computing the corresponding intensity distribution from the training set. These values were used to adjust the intensities in all datasets, including for training, validation, and testing. Intensities were then scaled so that the voxel values ranged between 0 and 1. Both classes in the training set were augmented ten-fold using random rotations, translations, scaling, horizontal flips, vertical flips, and Gaussian noise.

\subsection{Deep Learning Workflow}

This section outlines the process of selecting and optimizing model architectures for detecting Bankart lesions. 

We did not pool standard MRIs and MRAs into a single training set as the two modalities differed significantly in patient demographics and lesion prevalence (see \autoref{tab:demographics}), so mixing them without harmonization would risk the model learning modality or population-specific cues rather than pathology. We therefore trained and evaluated separate models per modality to avoid confounding. 

We leveraged pretraining on the MRNet dataset \cite{bien_deep-learning-assisted_2018}, a publicly available knee MRI dataset, to address the challenges posed by our small, imbalanced dataset. We also used pretrained weights from the ImageNet\cite{deng_imagenet_2009} dataset for comparison. Model selection began with hyperparameter tuning across various architectures. Top-performing models underwent additional optimization on our ScopeMRI shoulder dataset (\autoref{hparam_tuning}).

\begin{figure} [ht]
\begin{center}
\begin{subfigure}{\textwidth}
    \centering
    \includegraphics[width=\textwidth]{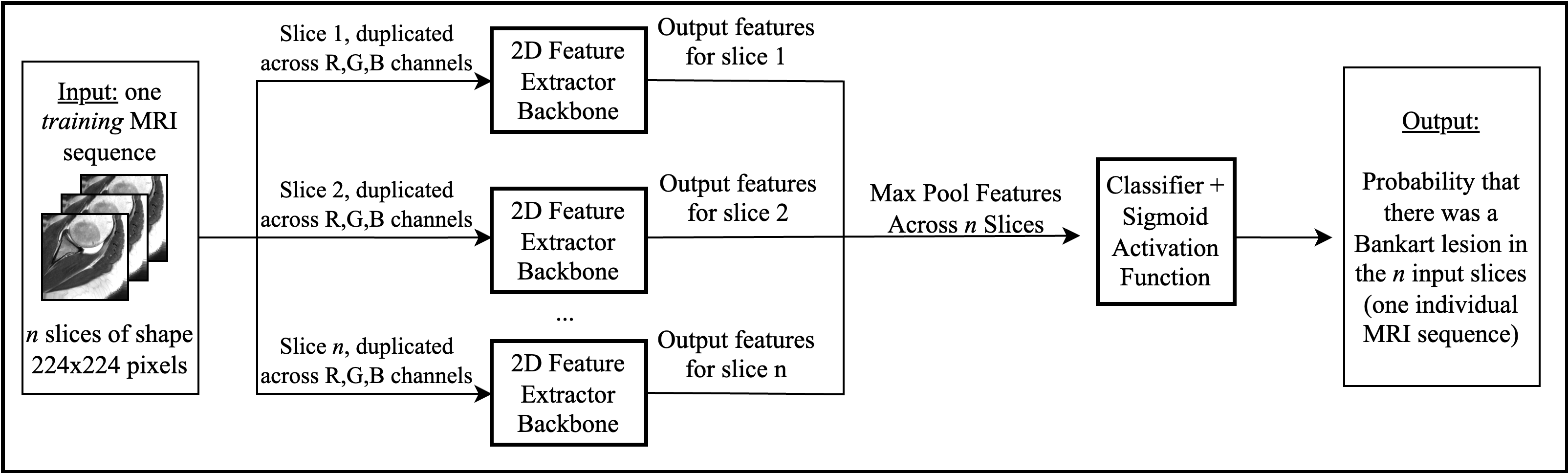}
    \caption{Schematic of 2D model training setup using 3D MRIs.}
    \label{subfig:train}
\end{subfigure}
\\  
\begin{subfigure}{\textwidth}
    \centering
    \includegraphics[width=\textwidth]{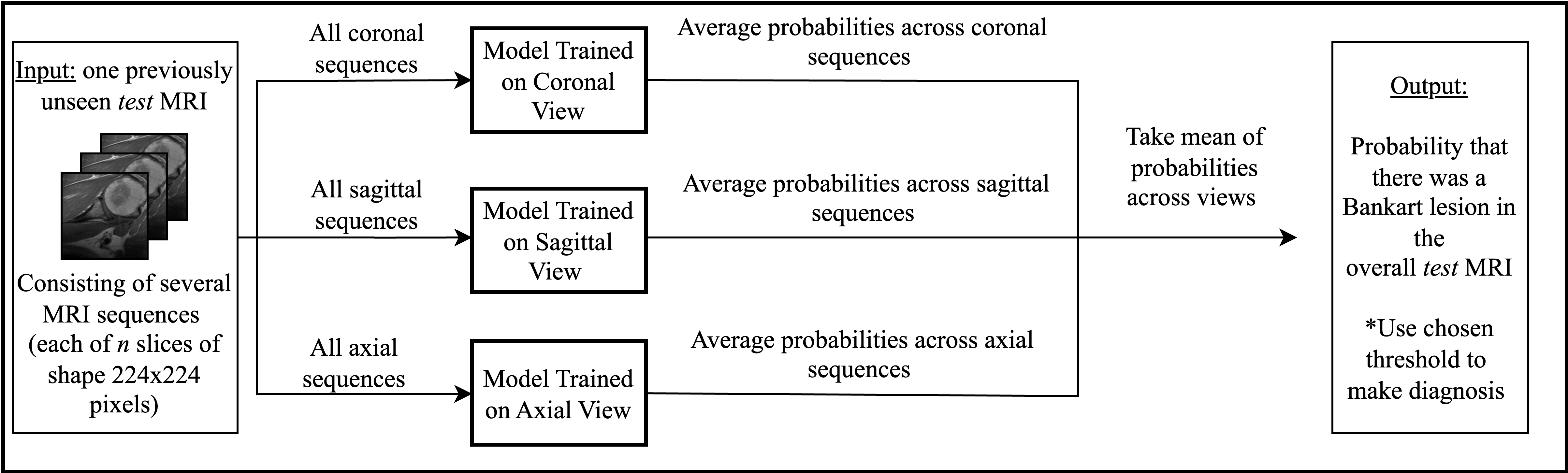}
    \caption{Schematic of model multi-view ensemble for inference.}
    \label{subfig:inference}
\end{subfigure}
\end{center}
\caption[Model Training \& Inference]
{\label{fig:train_eval} Model Training \& Inference. (A) Schematic of 2D model training setup using 3D MRIs. (B) Schematic of model multi-view ensemble for inference.
The setup for the 3D CNN only differed in that the entire preprocessed MRI volume was input directly into the model, then the output was fed into the classifier \& sigmoid layers.}
\end{figure}

To ensure stable model selection, we employed stratified cross-validation and prioritized stability across splits (\autoref{cv_info}). Final models were re-trained on the initial dataset split for sagittal, coronal, and axial views, which were combined into a multi-view ensemble for evaluation on a hold-out test set and an external dataset (\autoref{retrain_ensemble}). Performance metrics included accuracy, sensitivity, specificity, and area under the receiver operating characteristic curve (AUC). Post-hoc interpretability was assessed using Grad-CAM\cite{selvaraju_grad-cam_2017}, providing visual insights into model predictions (\autoref{methods_interpretability}). Implementation details and system requirements are described in \autoref{app:train_inf} and \autoref{app:system_req}, respectively. \autoref{fig:train_eval} contains a schematic overview of our training and inference setup. 

\subsection{Model Architecture \& Pretraining}
\label{model_architecture}

We evaluated multiple state-of-the-art architectures to address the diagnostic challenges of Bankart lesion detection. These included AlexNet\cite{krizhevsky_imagenet_2012}, DenseNet\cite{huang_densely_2018}, EfficientNet\cite{tan_efficientnet_2020}, ResNet34\cite{he_deep_2015}, ResNet50\cite{he_deep_2015}, Vision Transformer (ViT)\cite{dosovitskiy_image_2021}, Swin Transformer V1\cite{liu_swin_2021}, and Swin Transformer V2\cite{liu_swin_2022}, all initialized with ImageNet\cite{deng_imagenet_2009} weights. For each MRI, individual slices were passed through the 2D model feature extractor, and features were aggregated across slices using max pooling, resulting in vector $v$. Vector $v$ was then fed into a classifier layer, followed by a sigmoid function, to output a probability for each scan. Probabilities from MRI sequences from the same overall MRI were averaged together to get probabilities for each view. See \autoref{fig:train_eval} for additional details. 

We also evaluated a custom 3D CNN. The custom 3D CNN architecture was inspired by AlexNet due to its strong performance in the detection of knee pathologies on MRIs\cite{bien_deep-learning-assisted_2018}. It consisted of five convolutional layers with ${96, 256, 384, 384, 256}$ filters, followed by ReLU activations, batch normalization, and max pooling after the first, second, and fifth convolutional layers. An adaptive average pooling layer and two fully connected layers ($4096$ neurons each) were included in the classifier, with dropout applied before the final layer.

For pretraining, we used the MRNet dataset \cite{bien_deep-learning-assisted_2018}, a publicly available knee MRI dataset chosen for its anatomical and contextual relevance to musculoskeletal imaging. Pretraining for the initial architecture search was conducted on the sagittal view of MRNet using the \enquote{abnormal} label as the target. The top-performing architectures—AlexNet, Vision Transformer, and Swin Transformer V1—were identified based on their AUC on the MRNet validation set. These models were subsequently pretrained on the axial and coronal views of MRNet to prepare for fine-tuning on our shoulder dataset. Additional details on MRNet pretraining can be found in \autoref{app:pretrain}. 

\subsection{Hyperparameter Tuning}
\label{hparam_tuning}

Hyperparameter tuning was conducted separately for standard MRIs and MRAs using Hyperband optimization \cite{li_hyperband_2018}. The search space included learning rate (log-uniform distribution: 1e-8 to 1e-1), weight decay (log-uniform distribution: 1e-6 to 1e-1), dropout rate (uniform distribution: 0 to 0.5), and learning rate scheduler (choice between CosineAnnealingLR with $T_{max}$=10 and ReduceLROnPlateau with factor=0.5 \& patience=3).

To assess if MRNet pretraining resulted in performance improvements compared to ImageNet-initialization\cite{krizhevsky_imagenet_2012}, for each view (sagittal, axial, coronal) and modality (standard MRI, MRA), two Hyperband hyperparameter tuning runs were conducted: one where model weights were initialized via the MRNet pretraining strategy described in \autoref{model_architecture} and \autoref{app:pretrain}, and one using ImageNet-weights\cite{krizhevsky_imagenet_2012}. For both cases, weights for the classifier were randomly initialized. The hyperparameter combination achieving the highest AUC on our dataset's validation set was designated as the \enquote{best trial}. The hyperparameters and weight initialization strategy from this \enquote{best trial} were then applied to the stratified 8-fold cross-validation on our dataset. 

\subsection{Dataset Split \& Cross-Validation}
\label{cv_info}
Our dataset was initially split into a 20\% hold-out test set, as well as an 80\% training-validation (70\% training \& 10\% validation) set using random stratified sampling to preserve class ratios. The initial training-validation split was used for all hyperparameter tuning. \par

All splitting was performed at the MRI (shoulder) level. Each shoulder contributed at most one standard MRI and/or one MRA; if repeat imaging was obtained due to poor quality, the initial scan was excluded. Patients with prior ipsilateral shoulder surgery or prior injuries were excluded. In rare cases where a patient underwent both a standard MRI and an MRA of the same shoulder, both scans were retained, but the standard MRI and MRA datasets were modeled and evaluated separately. This design ensured that all sequences/views from a given shoulder remained within a single split (train, validation, or test), preventing data leakage.

Given the variability inherent in small, imbalanced datasets, we conducted cross-validation to evaluate model stability across different training-validation splits. This was achieved by recombining the original 10\% validation and 70\% training sets (in total, comprising 80\% of the original data), then using random stratified sampling to generate 8 folds. The cross-validation was performed using the hyperparameters identified from the best trial in the tuning phase described in \autoref{hparam_tuning}. For each view-modality pair, models were trained on seven folds and validated on the remaining fold. Validation AUC was recorded for each fold, and the architecture exhibiting the highest stability (lowest standard deviation in validation AUC) was selected for each view-modality combination.

\subsection{Re-Training on Initial Split and Multi-View Ensembling}
\label{retrain_ensemble}
To minimize potential biases introduced by selecting specific cross-validation splits, we re-trained the most stable model for each view-modality pair on the initial random training-validation split. This re-training also ensured consistency in the validation set across views so it could be used for prediction threshold selection. 

Probabilities from the coronal, sagittal, and axial view models were averaged to generate a multi-view ensemble prediction for both MRAs and standard MRIs. The prediction threshold for accuracy, sensitivity, and specificity was selected based on the point where sensitivity and specificity were equal on the unified validation set. This method was chosen to balance avoiding missed diagnoses while minimizing overcalls. The multi-view ensemble model was then evaluated on the hold-out test set and small external validation set, providing the final performance metrics for Bankart lesion detection. 

\subsection{Interpretability}
\label{methods_interpretability}
Grad-CAM \cite{selvaraju_grad-cam_2017} was used as a post-hoc interpretability tool to visualize model attention and identify regions within each image that contributed most to the model’s predictions. For convolutional architectures (i.e. AlexNet), the final convolutional layer was used to compute Grad-CAM outputs. For transformer-based architectures (i.e. Vision Transformer and Swin Transformer), the outputs from the final attention block were used.

Heatmaps were generated for representative examples from the axial view for both modalities (standard MRI \& MRA) for both positive (Bankart lesion present) and negative cases. These heatmaps were scaled to the original preprocessed image dimensions and overlaid on the corresponding MRI slices for visualization. This allowed for qualitative assessment of model attention alignment with clinically relevant features.




\acknowledgments 
 We would like to thank the University of Chicago Center for Research Informatics (CRI) High-Performance Computing team for providing resources and support throughout this project. We would also like to thank the University of Chicago CRI Clinical Research Data Warehouse \& Human Imaging Research Office (HIRO) for their roles in data collection. Finally, we would like to thank Steven Song and the UChicago AI in Biomedicine Journal Club for their constructive feedback on our manuscript. 

 The CRI is funded by the Biological Sciences Division at the University of Chicago with additional funding provided by the Institute for Translational Medicine, CTSA grant number UL1 TR000430 from the National Institutes of Health.

 This project was supported by the National Center for Advancing Translational Sciences (NCATS) of the National Institutes of Health (NIH) through Grant Number UL1TR002389-07 that funds the Institute for Translational Medicine (ITM). The content is solely the responsibility of the authors and does not necessarily represent the official views of the NIH.

 M.S. was funded by the U.S. Department of Energy, Office of Science, Office of Advanced Scientific Computing Research, Department of Energy Computational Science Graduate Fellowship under Award Number DE-SC0023112.

\section*{DECLARATIONS}
\noindent \textbf{Conflict of interest/Competing interests: }

\noindent The authors declare no conflicts of interest. 

\noindent \textbf{Ethics approval and consent to participate:} 

\noindent This study received approval from the University of Chicago's Institutional Review Board (IRB24-0025). 

\noindent \textbf{Data availability:}

\noindent SCOPE-MRI has been deposited to \url{https://www.midrc.org/}. Exact access instructions are found in the code repository. Due to Institutional Review Board (IRB) restrictions, data collected from external institutions cannot be publicly released, and therefore, the external dataset will not be included.

\noindent \textbf{Code availability:} 

\noindent To support reproducibility and broader applicability, we publicly release our training and evaluation code as a modular framework for binary classification on MRI and CT scans. The codebase includes support for training, hyperparameter tuning, cross-validation, and Grad-CAM visualizations. While originally developed for Bankart lesion detection, it is designed to apply deep learning to any binary classification task on MRIs or CTs. It is available at: \url{https://github.com/sahilsethi0105/scope-mri/}

\bibliography{references} 
\bibliographystyle{spiebib} 

\newpage
\label{sec:sup_info}
\begin{center}
{\LARGE \textbf{Supplementary Information}}\\[1ex]
{\large SCOPE-MRI: Bankart Lesion Detection as a Case Study in Data Curation and Deep Learning for Challenging Diagnoses}\\[1ex]
Sahil Sethi, Sai Reddy, Mansi Sakarvadia, Jordan Serotte, Darlington Nwaudo, Nicholas Maassen, \& Lewis Shi
\end{center}

\renewcommand\thesection{S\arabic{section}}
\renewcommand\thefigure{S\arabic{figure}}
\renewcommand\thetable{S\arabic{table}}
\renewcommand\theequation{S\arabic{equation}}

\setcounter{section}{0}
\setcounter{figure}{0}
\setcounter{table}{0}
\setcounter{equation}{0}

\section{Implementation Details}
\label{app:train_inf}

For all training, we used binary cross-entropy loss, with the loss weighted inversely proportional to class prevalence to address class imbalance. Early stopping was applied if validation accuracy did not improve for 10 consecutive epochs. During hyperparameter tuning, models were trained for a maximum of 20 epochs, with weights from the epoch achieving the highest validation accuracy retained for inference. For cross-validation, models were trained for up to 30 epochs. During final retraining on the full dataset, the maximum number of epochs was set to 100 to allow for potential further improvements, but early stopping was consistently applied. For all models, a batch size of 1 was used, as each batch contained, on average, 20–30 slices corresponding to a single MRI sequence. 

\section{Additional Context on Design Decisions}
\label{app:design_decisions}
Given the dataset scale and class imbalance, we expected slice-based models using 2D backbones pretrained on ImageNet to be more successful than 3D models; we nevertheless evaluated a custom 3D CNN in the initial search, which underperformed the top 2D candidates and was not advanced (see \autoref{model_selection_extraresults_section}). For the 2D backbones that were advanced as candidates, selection was performed empirically per view–modality under a fixed training pipeline because cross-validation consistently favored different backbones across views; while a uniform backbone across views could reduce architectural heterogeneity, the observed per-view differences suggest such a choice would likely trade some accuracy for simplicity. Next, for multi-view fusion, we averaged axial, sagittal, and coronal predictions equally to avoid introducing view-weighting hyperparameters that a small, imbalanced dataset cannot reliably estimate; more adaptive view-weighting would be interesting to explore in larger cohorts. Future work could explore learned view/sequence fusion or custom multi-view architectures, which could potentially improve performance.

\section{MRNet Pretraining}
\label{app:pretrain}
We evaluated the deep-learning architectures described in Section \enquote{Model Architecture and Pretraining} using pretraining on the MRNet dataset \cite{bien_deep-learning-assisted_2018}, specifically with the sagittal view and the \enquote{abnormal} label. The 120 MRI validation set from Bien et al. \cite{bien_deep-learning-assisted_2018} (called \enquote{tuning set} in their paper) was used as a hold-out test set, as the authors never publicly released their test set. For monitoring training, a new 120 MRI subset was selected and held aside from the training set using random stratified sampling (by label to maintain class ratios)—for clarity, we refer to this newly selected set as the \enquote{development} set. Each model underwent Hyperband optimization\cite{li_hyperband_2018} for 100 trials. Section \enquote{Hyperparameter Tuning} describes the hyperparameter tuning process on our dataset; the same process was used for MRNet pretraining except for the following changes: both classes were augmented five-fold rather than ten-fold, $T_{max}$=5 instead of 10 for CosineAnnealingLR, and the objective was to optimize MRNet development set AUC. 

For each architecture, the trial achieving the highest development set AUC proceeded to inference on the MRNet validation set (called \enquote{tuning} set in their paper). The three architectures showing the highest MRNet validation set AUC were selected for fine-tuning on our dataset. As these were trained on only the sagittal view, axial and coronal models were next trained on the MRNet dataset using the same process. Weights from these models trained on each view of MRNet were used for initializing the corresponding \enquote{MRNet-initialized} trials described in Section \enquote{Hyperparameter Tuning.}

\section{Initial Candidate Architecture Selection}
\label{model_selection_extraresults_section}

\autoref{tab:model_comparison} summarizes the performance of various architectures pretrained on the MRNet dataset. AlexNet, Vision Transformer (ViT), and Swin Transformer V1 achieved the highest AUC on the MRNet validation (\enquote{tuning}) set, and thus were selected for fine-tuning on our dataset. 

\begin{table}[H]
\centering
\caption{Results of Pretraining Various Architectures on MRNet \enquote{Abnormal} Label for Sagittal View}
\renewcommand{\arraystretch}{1.2}  
\newcolumntype{P}[1]{>{\centering\arraybackslash}p{#1}}
\begin{tabular}{p{0.5\linewidth} P{0.2\linewidth}}
\toprule
\textbf{Model Type}& \textbf{AUC on MRNet Validation Set}\\
\midrule
AlexNet \cite{krizhevsky_imagenet_2012} & \textbf{0.9242}\\
EfficientNet \cite{tan_efficientnet_2020} & 0.6472\\
DenseNet \cite{huang_densely_2018} & 0.4072\\
ResNet34 \cite{he_deep_2015} & 0.5848\\
ResNet50 \cite{he_deep_2015} & 0.7371\\
3D CNN (custom) & 0.7439\\
Vision Transformer (ViT) \cite{dosovitskiy_image_2021} & \textbf{0.9561}\\
Swin Transformer V1 \cite{liu_swin_2021}& \textbf{	
0.9465}\\
 Swin Transformer V2 \cite{liu_swin_2022}&0.9061\\
\bottomrule

\end{tabular}

\vspace{5pt} 
\parbox{0.85\textwidth}{%
\footnotesize%
}
\label{tab:model_comparison}
\end{table}

\section{Final Hyperparameters}
\label{hparam_tuning_extraresults_section}

\autoref{tab:model_hp} presents the optimal hyperparameters for each view-modality, identified through Hyperband\cite{li_hyperband_2018} tuning followed by cross-validation 
 (for identifying the most stable architecture). These hyperparameters were used for final model re-training on the initial dataset split.

\FloatBarrier
\begin{table}[H]
\centering
\caption{Final Hyperparameters for Each View-Modality}
\renewcommand{\arraystretch}{1}  
\newcolumntype{P}[1]{>{\centering\arraybackslash}p{#1}}
\begin{tabular}{p{0.225\linewidth}>{\raggedright\arraybackslash}p{0.1\linewidth}>{\raggedright\arraybackslash}p{0.1\linewidth}>{\raggedright\arraybackslash}p{0.175\linewidth}>{\raggedright\arraybackslash}p{0.075\linewidth}>{\raggedright\arraybackslash}p{0.075\linewidth}>{\raggedright\arraybackslash}p{0.075\linewidth}}
\toprule
\textbf{View-Modality}& \textbf{Model Type}&  \textbf{Weight Initialization}&\textbf{Learning Rate Scheduler (LR)}& \textbf{Initial LR}& \textbf{Weight Decay}&\textbf{Dropout Rate}\\
\midrule
Sagittal-MRA& ViT \cite{dosovitskiy_image_2021}&  MRNet&ReduceLROnPlateau& 5.40e-6& 9.98e-6&0.220\\
Axial-MRA& AlexNet \cite{krizhevsky_imagenet_2012}&  MRNet&ReduceLROnPlateau& 2.24e-5& 4.90e-4&0.303\\
Coronal-MRA& ViT \cite{dosovitskiy_image_2021}&  MRNet&ReduceLROnPlateau& 4.42e-6& 2.01e-4&0.114\\
Sagittal-Standard MRI& Swin V1 \cite{liu_swin_2021}&  MRNet&CosineAnnealingLR& 6.81e-6& 2.34e-3&0.261\\
Axial-Standard MRI& ViT \cite{dosovitskiy_image_2021}&  MRNet&CosineAnnealingLR& 5.82e-6& 1.45e-5&0.227\\
Coronal-Standard MRI& Swin V1 \cite{liu_swin_2021}&  MRNet&CosineAnnealingLR& 1.60e-6& 4.47e-4&0.059\\
\bottomrule

\end{tabular}

\vspace{5pt} 
\parbox{0.85\textwidth}{%
\footnotesize%
}
\label{tab:model_hp}
\end{table}
\FloatBarrier

\section{System Requirements}
\label{app:system_req}
 All model training and inference was performed on one NVIDIA A100-40GB GPU. Models were implemented using Python (version 3.10) and the PyTorch library (version 2.3.1). 



\end{document}